\documentclass[12pt]{article}
\usepackage{graphicx}
\def\beq{\begin{equation}}
\def\eeq{\end{equation}}
\def\bea{\begin{eqnarray}}
\def\eea{\end{eqnarray}}

\def\roughly#1{\mathrel{\raise.3ex\hbox
{$#1$\kern-.75em\lower1ex\hbox{$\sim$}}}}

\def\sss{\scriptscriptstyle}

\def\Yud{Y^{U,D}}
\def\Sud{S^{U,D}}

\def\barpk{{\raise.35ex\hbox  {${\sss  (}$}}--{\raise.35ex\hbox{${\sss
)}$}}}        \def\bbarp{\hbox{$B$\kern-0.9em\raise1.4ex\hbox{\barpk}}}

  \def\rr2{{1\over\sqrt{2}}}

\def\.{\!\cdot\!}    \def\:{\cdots}   \def\[{\left[}   \def\]{\right]}
\def\({\left(} \def\){\right)} 
\def\Journal#1#2#3#4{{#1} {\bf #2}, #3 (#4)}

\def\NPB{{\em Nucl. Phys.} B}
\def\PLB{{\em Phys. Lett.}  B}
\def\PRL{\em Phys. Rev. Lett.}
\def\PRD{{\em Phys. Rev.} D}

\def\rr2{{1\over\sqrt{2}}}
\begin{document}
\begin{flushright}
UMISS-HEP-2008-04 \\
[10mm]
\end{flushright}

\begin{center}
\bigskip {\Large  \bf Suppressed FCNC in New Physics with Shared  Flavor Symmetry}
\\[8mm]
Alakabha Datta\footnote{E-mail:
\texttt{datta@phy.olemiss.edu}}
\\[3mm]
\end{center}

\begin{center}
~~~~~~~~~~~ {\it Department  of Physics and Astronomy,}\\ 
~~~~~~~~~~~~{ \it University of Mississippi,}\\
~~~~~~~~~~~~{\it  Lewis Hall, University, MS, 38677.}\\
\end{center}

\newpage

\begin{center} 
\bigskip (\today) \vskip0.5cm {\Large Abstract\\} \vskip3truemm
\parbox[t]{\textwidth} {  
Many extensions of the Standard Model(SM) generate contributions to Flavor Changing Neutral Current(FCNC) processes that must have sufficient flavor suppression to be consistent with experiments, if the new physics (NP) is associated with a scale of a TeV.  
Here we present a   mechanism  for suppressing the NP effects to FCNC processes. We consider the possibility that the source of NP contributions to FCNC processes  share the same flavor symmetry underlying the SM source of FCNC processes which are the quark and lepton mass matrices. We call this the principle of shared flavor symmetry. In the flavor symmetric limit, the quark and lepton mixing matrices  have  fixed forms and there  are no NP FCNC processes. In the flavor symmetric limit, we take the quark mixing matrix to be the identity matrix and the lepton mixing matrix to be given by tri-bimaximal mixing. Realistic mixing matrices are obtained by the small breaking of the flavor symmetry. New contributions to FCNC processes arise because of non universal breaking of the flavor symmetry in the quark and lepton mass matrices and the NP sources of FCNC processes. In particular, we will  focus on  new FCNC effects that  arise due to the breaking of flavor symmetry only in the quark and charged lepton mass matrices but not in the NP sector. In this scenario,
NP contributions to FCNC processes is linked to the source of  flavor symmetry breaking in the quark and charged lepton mass matrices. 
 The breaking of flavor symmetry in the NP sector is assumed to produce FCNC effects that are  at most the size of  NP FCNC effects due to the breaking of flavor symmetry in the quark and charged lepton mass matrices.  
To demonstrate the mechanism we use a two higgs doublet model as an example of beyond the SM physics though  one should be able to adapt this mechanism to other models of new physics. 
  }
\end{center}

\thispagestyle{empty} \newpage \setcounter{page}{1}
% Decrease texheight (for preprint numbers) again
%\textheight 23.0 true cm
\baselineskip=14pt

\section{Introduction}
Flavor Changing Neutral Current (FCNC) processes in the Standard Model(SM) do not arise at tree level and are highly suppressed. Many extensions of the SM naturally have FCNC processes that occur at tree  or loop level. These processes, if they involve new particles  of mass $\sim $ TeV, must be sufficiently suppressed to be consistent with experiments. In the quark sector, in the SM, there are FCNC processes where the internal quarks in the loops are the third family quarks. These FCNC processes are not only loop suppressed but are also flavor suppressed by various powers of $\lambda$
where $\lambda \sim 0.23$ is the Cabibbo angle. In the down type quark sector, the $b \to s$ transition is flavor suppressed by $\sim \lambda^2$, the $b \to d$ transition is flavor suppressed by $\sim \lambda^3$ and the $s \to d$ transition is flavor suppressed by $\sim \lambda^5$. New physics (NP)  contributions to these FCNC processes involving TeV scale particles should have at least the same flavor suppressions to produce effects similar in size to the SM. There must be a mechanism to produce
such flavor suppressions in the NP contributions to the FCNC processes.
 %to agree with measurements. 
 In the top quark sector, the $t \to c$ transition is suppressed by $\sim \lambda^2$, the $t \to u $ transition is suppressed by $\sim \lambda^3$ and the $c \to u$ transition is suppressed by $\sim \lambda^5$. These processes have an additional suppression due to the small b quark mass compared to the W boson mass.  
There are also FCNC contributions in the SM which have milder flavor suppression but are tiny due to small internal quark masses. As an example the $s \to d$ transition with an internal charm quark only has a flavor suppression of $\sim \lambda$. Any NP with the same flavor suppression and TeV scale particles will in general produce very large effects in conflict with experiments.
Hence there must also be a mechanism to forbid such NP contributions.

 In the leptonic sector experiments have imposed stringent limits on FCNC processes \cite{fcnclepton}. This implies that NP contributions to FCNC in the leptonic sector must also have significant flavor suppression for TeV scale NP. Since mixing in the leptonic sector is  mostly large
 there is little flavor suppression in the FCNC generated though loops involving SM particles. In the SM, the tiny internal lepton
masses in the loops are responsible for  the extremely tiny size of FCNC effects. A mechanism  to produce large flavor suppression in FCNC is then necessary in NP with TeV scale particles.

To be specific let us choose a specific model of new physics with two higgs doublets and let us concentrate on FCNC in the quark sector for the moment. One of the higgs doublet, $\phi_1$ couples to the quarks with Yukawa couplings $Y^{U,D}$ while the second doublet , $ \phi_2$, couples to the quarks with Yukawa couplings $S^{U,D}$.  As one transforms the quarks to the mass basis by diagonalizing, $Y^{U,D}$, there are FCNC effects associated with the neutral component of the second higgs doublet arising from the non-diagonal Yukawa couplings in
$S^{U,D}$. The natural size of these non diagonal couplings are $O(1)$ and hence, 
to be consistent with experiments the mass of the second higgs doublet must  be very high, much greater than a TeV. Now it is widely believed that there must be NP around a TeV to account for the stability of the SM higgs mass and hence FCNC in NP must be sufficiently suppressed to allow NP to be in the TeV region.

A popular suggestion that provides suppression of new FCNC effects is the principle of minimum flavor violation (MFV) \cite{MFV}. In this scheme the CKM matrix is the only source of flavor mixing and CP violation even with NP. We can describe the principle by using the two higgs doublet model described above. In the SM, the Yukawa interactions break a flavor symmetry $ H \equiv SU(3)_Q \times SU(3)_U \times SU(3)_D$ where the various $SU(3)$ factors act on the left handed quark doublets, the right handed up type quarks and the right handed down type quarks. Using spurion methods one can imagine the SM Yukawa couplings $Y^{U,D}$ to transform appropriately under $H$ to make the Yukawa interactions invariant under $H$. MFV requires the new Yukawa couplings 
$S^{U,D}$ to have the same transformation under $H$ as  $Y^{U,D}$. To implement this one can assume $S^{U,D}$ to be proportional to  $Y^{U,D}$ in which case there are no new tree level FCNC contributions or one can construct the $S^{U,D}$ out of the $Y^{U,D}$ and on going to the quark mass basis tree level FCNC associated with the second higgs doublet is suppressed and dependent on the CKM matrix \cite{wise}.

In this paper we present a different scheme to suppress new FCNC effects
which we will call the shared flavor symmetry (SFS) principle. Again we explain this principle with the two higgs doublet model of NP though this principle should be adaptable to other extensions of the SM.
We consider the possibility that the Yukawa couplings $\Yud$ and $\Sud$ share a flavor symmetry. In the flavor symmetric limit diagonalization of $\Yud$ automatically leads to the diagonalization of the matrices $\Sud$ and so there are no new FCNC. The transformation matrices diagonalizing $\Yud$ and $\Sud$ are pure numbers and are independent of the elements of $\Yud$ and $\Sud$. The CKM matrix  is the identity matrix in the flavor symmetric limit and there are no flavor changing and hence no FCNC processes in the full theory. 
The realistic CKM matrix is obtained by the breaking of the flavor symmetry. However the breaking of the flavor symmetry is non universal and is different for $\Yud$ and $\Sud$. This leads to tree level FCNC effects associated with the non diagonal elements of $\Sud$. A possible scenario is flavor symmetry breaking in only $\Yud$ and not in $\Sud$. In this case all NP FCNC effects are determined solely by the structure of the SM Yukawa couplings $\Yud$. In this paper we will mainly focus on this scenario.
 
Generally, there are two sources of new contributions to FCNC processes. One is from the breaking of flavor symmetry in $\Yud$ and the other from the breaking of flavor symmetry in $\Sud$. We can obtain information on the breaking of flavor symmetry in $\Yud$ from the quark masses and mixing. The breaking of flavor symmetry in $\Sud$ is unknown.
We will consider some scenarios of flavor symmetry breaking in $\Sud$. It is possible that the dominant new FCNC contributions arise from the breaking of flavor symmetry in $\Yud$.
 We will assume that FCNC effects due to the breaking of flavor symmetry in $\Sud$ are smaller or at most the same size as the FCNC effects from the flavor symmetry breaking in $\Yud$. 
 %The study of the source of flavor symmetry breaking in $\Yud$ and %$\Sud$ is beyond the scope of the paper. 

The flavor symmetry breaking in $\Yud$ in turn leads to the realistic CKM matrix and its deviation from the identity matrix. It is a natural assumption to take the flavor symmetry breaking in $\Yud$ to follow the same pattern as in the  CKM matrix. 
Hence  NP FCNC effects are expected to exhibit the following pattern: FCNC effects involving the second and third generation is suppressed by $ \sim \lambda^2$, while NP FCNC effects involving the first and third generation is suppressed by $ \sim \lambda^3$.  There is obviously a problem for FCNC involving the first and the second generation which should be suppressed by $ \lambda$. This suppression is clearly inconsistent with measurements in the neutral kaon and the $D$ meson system if the NP scale is a TeV.

To see how the principle of SFS may solve this problem, note that in the SM, with mixing only between the first two generations, all FCNC vanish if the higgs coupling to the quarks are the same for the two generations in the mass basis. This is the same as requiring $m_u=m_c$ and $m_d=m_s$.
 This suggests we can consider a flavor symmetric limit where
the higgs coupling to the quarks are the same for the two generations in the mass basis. According to the principle of SFS this symmetry is also shared by $\Sud$. Hence the second higgs also has the same coupling to the first two generations in the mass basis in the flavor symmetric limit.

This flavor symmetry is obviously broken in $\Yud$ but we will assume it remains intact in $\Sud$. With this assumption, we will be able to show in our NP model that  FCNC involving the first and second generation is suppressed by $ \sim \lambda^5$. 
%In fact all FCNC vanish when mixing involving the third generation is %turned off. 
%In this limit,
 %if we include only mixing between the first two generations via a %unitary matrix, the Yukawa matrices $\Sud$ will still continue to be    % diagonal in the mass basis and hence there are no new FCNC %contributions.
 {} For the mass matrices in our model, equal coupling of the second higgs to the quarks of the first two generations
 can be
 achieved by simply requiring that all NP FCNC effects vanish when the symmetry breaking involving the second and third generation is turned off.  

Let us now turn to the lepton sector. Leptonic mixing is almost always large and one should expect  large new FCNC effects  according to SFS. But this assumes that  the leptonic mixing is the identity matrix in the flavor symmetric limit. Current experimental data on leptonic mixing is consistent with what is known as the tri-bimaximal mixing \cite{TBM}. We assume that the tri-bimaximal mixing is the flavor symmetric limit of leptonic mixing.
Hence according to SFS all NP FCNC effects in the lepton sector depend on flavor symmetry breaking that cause deviations from the tri-bimaximal mixing. These deviations are small( see Ref.~\cite{dattaTBM}) and may have some similarities with flavor suppressions  in the quark sector.

The paper is organized in the following manner
We begin in Sec. 2 with a discussion on the shared flavor symmetry (SFS) principle. We  demonstrate the principle though a   
  two higgs doublet model extension of the SM. In 
Sec. 3 we study the effect of flavor symmetry breaking to generate the realistic CKM matrix and new FCNC effects. In the leptonic sector we consider flavor symmetry breaking in the charged lepton sector. This leads to deviations from the tri-bimaximal mixing and new FCNC effects involving the charged leptons. Finally, in Sec. 4 we present our conclusions.
%%%%%%%%%%%%%%%%%%%%%%%%%%%%%%%%%%%%%%%%%%%%%%%%%%%%%%%%%%%%%%%%%%%%%%%%

\section{ Shared Flavor Symmetry Principle}
In this section we discuss the suppression of FCNC effects with NP via the principle of shared flavor symmetry. We use  a
two  higgs  doublet model  (2HDM) to explain the principle.
 %The SPS may be applied to the charged lepton sector but the unknown %nature of neutrino masses and mixing makes predictions in the lepton %sector somewhat difficult  and hence we will limit our discussion of %our model to the quark sector only.
 
 We will write the Lagrangian responsible for masses and FCNC processes
 as,
 \bea  
 {\cal  L}=  {\cal  L}^{Q} +  {\cal  L}^{L},\
 \eea
where 
\bea  
 {\cal  L}^{Q}& = &  \xi_U Y^{U}_{ij} \bar  Q_{i,L}  \tilde\phi_1
U_{j,R}  +   \xi_D Y^D_{ij}  \bar  Q_{i,L}\phi_1 D_{j,R}  +  \zeta_U S^{U}_{ij}  \bar
Q_{i,L}\tilde\phi_2 U_{j,R} + \zeta_D S^D_{ij}\bar Q_{i,L} \phi_2 D_{j,R} \,+\,
h.c.  ,\nonumber\\
{\cal  L}^{L}&_= &    \xi_L Y^L_{ij}  \bar  L_{i,L}\phi_1 E_{j,R}  +  \zeta_L S^L_{ij}\bar L_{i,L} \phi_2 E_{j,R} + {\cal L}_{\nu} \,+\,
h.c.  ,\
\label{lag1}
\eea
\noindent where $\phi_i$, for $i=1,2$, are the two scalar doublets of
a 2HDM,  while  $Y^{D,U,L}$ and $S^{D,U,L}$  are the non-diagonal
matrices of the Yukawa couplings. The Lagrangian ${\cal L}_{\nu}$ is responsible for neutrino masses through some mechanism that shall remain unspecified. We will consider FCNC in the lepton sector that arises from the Yukawa couplings.

For convenience  we can choose to  express $\phi_1$ and  $\phi_2$ in a
suitable basis  such that  only the $Y^{D,U,L}$  couplings generate
the fermion masses. In such a basis one can write \cite{soni},
  \beq 
  \langle\phi_1\rangle=\left(
\begin{array}[]{c}
0\\ {v/\sqrt{2}}

\end{array}
\right)\,\,\,\, , \,\,\,\, \langle\phi_2\rangle=0 \,\,\,.  \eeq
\noindent The two higgs doublets in this case are of the form,
\bea   
\phi_1   &=  &   \frac{1}{\sqrt{2}}\pmatrix{0   \cr  v+H^0}   +
 \frac{1}{\sqrt{2}}\pmatrix{ \sqrt{2} \chi^+ \cr i \chi^0}, \nonumber\\
 \phi_2 &= & \frac{1}{\sqrt{2}}\pmatrix{ \sqrt{2} H^+ \cr H^1+i H^2}. \
 \eea
 
In principle there  can be mixing among the neutral  higgs but here we
neglect such mixing for the moment as our focus is on the Yukawa couplings.
We assume  the doublet $\phi_1$  corresponds to the scalar  doublet of
the SM and  $H^0$ to the SM higgs field.

Let us assume that the same bi-unitary transformation diagonalizes both the Yukawa matrices $Y^{D,U,L}$ and $S^{D,U,L}$ to the diagonal forms $Y^{D,U,L}_{diag}$ and $S^{D,U,L}_{diag}$.
We can therefore write,
\bea
Y^{D,U,L} & = & U_{D,U,L} Y^{D,U,L}_{diag} V^{\dagger}_{D,U,L},\nonumber\\
S^{D,U,L} & = & U_{D,U,L} S^{D,U,L}_{diag} V^{\dagger}_{D,U,L}.\
\label{gp}
\eea
The entries in $Y^{D,U,L}_{diag}$ and $S^{D,U,L}_{diag}$ are assumed to be independent of each other. The matrices $U_{D,U,L}$ are unique up to a diagonal phase matrix which will be chosen such that the elements of $Y^{D,U,L}_{diag}$, the quark and charged lepton masses, are  real numbers. 
The  diagonal quark and lepton mass matrices are related to $Y^{D,U,L}_{diag}$, 
using Eq.~\ref {lag1}, as,
\bea
M^D_{diag} & = &\pmatrix{\pm m_d & 0
&  0  \cr  0  &  \pm m_s  &  0 \cr  0  & 0  & \pm m_b
} =
\xi_D \frac{v}{\sqrt{2}}\pmatrix{y_d & 0
&  0  \cr  0  &  y_s  &  0 \cr  0  & 0  & y_b
}, \nonumber\\
M^U_{diag} & = &\pmatrix{\pm m_u & 0
&  0  \cr  0  & \pm m_c  &  0 \cr  0  & 0  &  \pm m_t
} =
\xi_U \frac{v}{\sqrt{2}}\pmatrix{y_u & 0
&  0  \cr  0  &  y_c  &  0 \cr  0  & 0  & y_t
}, \nonumber\\
M^L_{diag} & = &\pmatrix{\pm m_e & 0
&  0  \cr  0  & \pm m_{\mu}  &  0 \cr  0  & 0  & \pm m_{\tau}
} =
\xi_L \frac{v}{\sqrt{2}}\pmatrix{l_e & 0
&  0  \cr  0  &  l_{\mu}  &  0 \cr  0  & 0  & l_{\tau}
}. \
\label{mass_Yukawa}
\eea

 The diagonal elements of $S^{D,U,L}_{diag}$ can be complex and there are no phase degrees of freedom left in $U_{D,U,L}$ to make the diagonal elements real.

To be specific let us concentrate on the down type quark sector. We can express the elements of $Y^D$ and $S^D$ as
\bea
{Y^D_{ij}} & = & \sum_k y_k \Theta^k_{ij},\nonumber\\
{S^D_{ij}} & = & \sum_k s_k \Theta^k_{ij}, \nonumber\\
\Theta^k_{ij} & = & (U_D)_{ik}(V^{\dagger}_D)_{kj}, \
\label{gpelement}
\eea
where $y_k$ and $s_k$ are the diagonal elements in $Y^D_{diag}$ and $S^D_{diag}$. 
According to the principle of shared flavor symmetry $\Theta^k_{ij}$
are pure numbers.

Now, $\Theta^k_{ij}$ being fixed numbers  implies that all the elements of $Y^D$ are linear combinations of the  $y_k$   values and the elements of $Y^D$  are specified by 3 independent real parameters. 
%Assuming the $y_k$ values to be non-degenerate, which is the case for %$Y^D$, we can solve for $y_k$ in terms of  three elements of $Y^D$ which %are different. All other elements of $Y^D$ are then expressible as %linear combination of these three elements . 
This at once implies some form of flavor symmetry as not all elements of $Y^D$ are independent. The matrix $S^D$ will share the same flavor symmetry but the elements of the matrix will be in general complex and different from that in $Y^D$. 
%The same conclusions apply to $S^D$. However, not all set of two %matrices specified by 3 different parameters will be simultaneously %diagonalizable.

It is instructive to consider a  specific examples to demonstrate the ideas above. Consider a structure for the Yukawa matrices which has a 2-3 flavor symmetry \cite{mutau}
\bea
Y^D & = &\pmatrix{y_{11} & y_{12}
&  -y_{12}  \cr  y_{12}  &  y_{22}  &  y_{23}\cr  -y_{12}  &y_{23}  &y_{22}
}. \
\eea
This matrix is diagonalized as
\bea
U^{\dagger} Y^D U & = & Y^D_{diag}, \nonumber\\
U & = &\pmatrix{1 & 0
&  0  \cr  0  & 1/\sqrt{2}   &  1/\sqrt{2}\cr  0  &- 1/\sqrt{2}  &1/\sqrt{2}
} 
\cdot
\pmatrix{\cos{\theta} & \sin{\theta}
&  0  \cr  -\sin{\theta}  & \cos{\theta}   &  0\cr  0  &0  &1
},\ 
\label{lam}
\eea
where the mixing angle $\theta$ is determined by the positive solution to
 \bea
 \tan \theta  & = & {2 \sqrt{2}y_{12} \over {
 y_{22}-y_{23}-y_{11} \pm 
 \sqrt{ (y_{22}-y_{23}-y_{11})^2+8y_{12}^2}
 }
 }. \
 \eea
 The eigenvalues of $Y^D$ are ${ 1 \over 2} [y_{11}+y_{22}-y_{23} \pm \sqrt{(y_{11}-y_{22}+y_{23})^2+8y_{12}^2} ]$ and $y_{22}+y_{23}$.
It is clear from the above that the transformation that diagonalizes $Y^D$  depends on the elements of $Y^D$ and so  the same transformation will not diagonalize the matrix $S^D$ which has the same structure as $Y^D$ but different matrix elements.

According to the SPS principle the elements of the matrix that diagonalizes the Yukawa matrices must be pure numbers. It is clear that we can achieve that
 by setting $y_{12}=0=s_{12}$( $\theta =0$) , and now both the Yukawa  matrices can be simultaneously diagonalized. We are going to consider this limit of the 2-3 symmetry, which we will call the decoupled 2-3 symmetry, as the flavor symmetry in the quark sector and the charged lepton sector. In this decoupled 2-3 symmetric limit \cite{model23} the first generation is decoupled from the second and third generations.
 
Let us represent the Yukawa couplings with the decoupled 2-3 symmetry by $Y^D_{23}$ and $S^D_{23}$.

We will write  the Yukawa couplings in the down quark sector as,
\bea
Y^D_{23} & = &
\pmatrix{y_{11} & 0
&  0  \cr  0  &  \frac{1}{2}{y_{22}}  &  \frac{1}{2}{y_{23}}\cr  0  
&\frac{1}{2}{y_{23}}  &\frac{1}{2}{y_{22}}}, 
 \nonumber\\
 S^D_{23} & = &
\pmatrix{q_{11} & 0
&  0  \cr  0  &  \frac{1}{2}{q_{22}}  &  \frac{1}{2}{q_{23}}\cr  0  & \frac{1}{2}{q_{23}}  & \frac{1}{2}{q_{22}}}. 
 \
 \label{23dsym}
 \eea
The Yukawa matrices  $Y^D_{23}$ and $S^D_{23}$ are diagonalized by the same unitary matrix $W^{d}_{23}$ given by,
\bea
W^d_{23} &  = & \pmatrix{1 & 0
&   0  \cr   0   &  -\frac{1}{\sqrt{2}}   &  \frac{1}{\sqrt{2}}\cr   0
&\frac{1}{\sqrt{2}} &  \frac{1}{\sqrt{2}}}.\
\label{wd} 
\eea 
Note that this matrix differs from the one in Eq.~\ref{lam} in the limit $\theta=0$ by an irrelevant diagonal phase matrix. 
%The fact that the CKM matrix is hernial does not necessarily mean %that the quark masses have to be hierarchical too.

Writing the diagonalized Yukawa matrices as $Y^{D}_{23diag}$ and
$S^D_{23diag}$ we have,
\bea 
Y^D_{23diag} & = & W^{d \dagger}_{23} Y^D_{23} W^d_{23} =\pmatrix{y_{11} & 0
&  0  \cr  0  &   \frac{1}{2}{(y_{22}-y_{23})}  &  0\cr  0  &0  &
\frac{1}{2}{(y_{22}+y_{23})}}, \nonumber\\ 
S^D_{23diag} & = & W^{d \dagger}_{23\dagger} S^D_{23} W^d_{23} =\pmatrix{q_{11} & 0
&  0  \cr  0  &   \frac{1}{2}{(q_{22}-q_{23})}  &  0\cr  0  &0  &
\frac{1}{2}{(q_{22}+q_{23})}}. \nonumber\\ 
\label{diag23sym} 
\eea 

%%%%%%%%%%%%%%%%%%%%%%%%%%%%%%%%%%%%%%%%%%%%%%%%%%%%%%%%%%%%%%%%
If we assume the same flavor symmetry  to apply to  the up quark sector then the Yukawa couplings, $Y_U$ and $S_U$,  are  diagonalized by the unitary matrix $W^u_{23}=W^d_{23}$. The CKM matrix $V_{CKM}=W^{u \dagger}_{23}W^d_{23} = I$. Hence there is no FC or FCNC processes in the theory.
We note that the flavor symmetry of the down and up quark sector do not have to be the same. However, if we require $V_{CKM}$ to be the identity matrix in the flavor symmetric limit then,
\bea
V_{CKM}^{sym} & = & V_{uL}^{\dagger}V_{dL}=I.\
 \label{ckm_sym}
 \eea
 where $V_{uL}$ and $V_{dL}$ transform the left handed up and down quarks from the gauge to the mass basis and $V_{CKM}^{sym}$ is the CKM matrix in the flavor symmetric limit.
 
 In the lepton sector we will also assume that there exists a flavor symmetric limit where the mixing matrix takes on an fixed form and has elements that are pure numbers. A possible form of this fixed MNS matrix is the tri-bimaximal form given as \cite{TBM},
 \bea
 U_{MNS}^{sym} &= &
\pmatrix{
\sqrt{\frac{2}{3}}  & \frac{1}{\sqrt{3}} & 0 \cr
-\frac{1}{\sqrt{6}}  & \frac{1}{\sqrt{3}} & \frac{1}{\sqrt{2}} \cr
\frac{1}{\sqrt{6}}  & -\frac{1}{\sqrt{3}} & \frac{1}{\sqrt{2}}
}.\
\label{tbm}
\eea
Any new FCNC processes in the lepton sector is then linked to deviation of the realistic leptonic mixing matrix from the TBM structure.

The TBM mixing is a specially case of mixing matrix with 2-3 flavor symmetry \cite{mutau}. In the 2-3 symmetry limit in the leptonic sector, 
the PMNS matrix, with $s_{13}=0$, is
given by,
\bea
U_{PMNS}^{s} & = & \pmatrix{c_{12} &s_{12}&0\cr
-\rr2 s_{12}
&\rr2 c_{12}  & \rr2\cr
\rr2 s_{12}
&- \rr2 c_{12} &
 \rr2}.
\label{mns23}
\eea
The TBM form is obtained by setting $s_{12}= { 1 \over {\sqrt{3}}}$.
As in Ref.~\cite{model23} we will assume a decoupled 2-3 symmetry in the charged lepton mass matrix.
We can then express $U_{PMNS}^{s}$, as,
\bea
U_{PMNS}^{s}&= U^{\dagger}_\ell U_\nu, \
\label{nulepton}
\eea
where
\bea
U^{\dagger}_\ell & = &
\pmatrix{1 & 0 &
0 \cr
0
& -\frac{1}{\sqrt{2}} & \frac{1}{\sqrt{2}}\cr
0
&\frac{1}{\sqrt{2}} &
 \frac{1}{\sqrt{2}}},\nonumber\\
U_\nu & = & \pmatrix{c_{12}&-s_{12}&0\cr s_{12}&c_{12}&0\cr 0&0&1\cr}\pmatrix{1&0&0\cr 0& -1&0\cr 0&0&1\cr}.\
\label{tbmfactor} 
\eea 
So in the flavor symmetric limit, the neutrino matrix, $U_{\nu}$ is just a combination of a simple rotation matrix and a 
phase matrix. 
We will consider breaking  of the decoupled 2-3 symmetry in the charged leptonic sector that will cause deviation from the TBM form and hence to new FCNC effects in the leptonic sector.

%%%%%%%%%%%%%%%%%%%%%%%%%%%%%%%%%%%%%%%%%%%%%%%%%%%%%%%%%%%%%%%%%%%%%%
 \section{Symmetry Breaking}
 We now consider symmetry breaking in the quark and the lepton sector. We will start with the quark sector and concentrate on the down type quark sector.
 FCNC effects in the down sector is specially interesting as there several hints of new physics FCNC contribution in certain rare B decays\cite{bnp}.
 
 Let us start with the decoupled 2-3 symmetric Yukawa couplings in Eq.~\ref{23dsym}. Using Eq.~\ref{diag23sym} we find that 
 the down  type quark mass matrix,  $M^D$ is  now given by,
\bea 
M^D_{diag} & = & W^{d \dagger}_{23} M^D W^d_{23} =\xi_D\pmatrix{\frac{v}{\sqrt{2}}y_{11} & 0
&  0  \cr  0  &   \frac{v}{\sqrt{2}}\frac{1}{2}{(y_{22}-y_{23})}  &  0\cr  0  &0  &
\frac{v}{\sqrt{2}}\frac{1}{2}{(y_{22}+y_{23})}}, \nonumber\\ 
W^d_{23} &  = & \pmatrix{1 & 0
&   0  \cr   0   &  -\frac{1}{\sqrt{2}}   &  \frac{1}{\sqrt{2}}\cr   0
&\frac{1}{\sqrt{2}} &  \frac{1}{\sqrt{2}}}.\ 
\eea

The  down type quark  masses   are  given   by,
  \bea  m_d   &  =   &  \pm
\xi_D\frac{v}{\sqrt{2}}y_{11},       \nonumber\\
  m_s       &=&      \pm
\xi_D\frac{v}{\sqrt{2}}{(y_{22}-y_{23}) \over 2},   \nonumber\\
  m_b   &  =   &   \pm
\xi_D \frac{v}{\sqrt{2}}{(y_{22}+y_{23}) \over 2}.\ 
\eea

Since $m_s <<  m_b$ there has to be a  fine tuned cancellation between
$y_{22}$ and $y_{23}$ to produce  the strange quark mass. Hence, it is
more  natural to  consider  the symmetry  limit $y_{22}=y_{23}$  which
leads to  $m_s=0$. The strange quark  mass is then generated due to symmetry breaking. We, therefore, consider the structure,
\begin{eqnarray}
 Y^D_{23} &= &\pmatrix{y_d & 0 &  0 \cr 0 & \frac{1}{2}{y_B}(1+2\chi_d) & \frac{1}{2}{y_B }\cr 0
&\frac{1}{2}{y_B} & \frac{1}{2}{y_B}}. \
 \label{23symbreak}
 \eea
Note that we  do not
break  the $2-3$ symmetry  in the  $23$ element  so that  the Yukawa
matrix remains  symmetric. This Yukawa matrix  is now diagonalized
by, $U_d= W^d_{23} R^d_{23}$ where
\bea
R^d_{23} & = & \pmatrix{1 & 0
&  0  \cr  0  &  c_{23d}   &  s_{23d}\cr  0  &-s_{23d}  &
c_{23d}}, \nonumber\\ 
c_{23d} & = & \cos { \theta_{23d}} ; s_{23d}  =  \sin { \theta_{23d}}. \
\label{23d}
 \eea
 We can write $Y^D_{23}$ as
\bea
Y^D_{23} & = & U_d \pmatrix{y_d & 0
&  0  \cr  0  &  y_{s}  &  0 \cr  0  & 0  & y_{b}
}U_d^{\dagger}, \
\label{another}
\eea
 where $y_{d,s,b}$ are the diagonal Yukawa couplings defined in Eq.~\ref{mass_Yukawa}.
One can then obtain 
 the non zero elements of $Y^D_{23}$ as,
\bea
(Y^{D}_{23})_{11} & = & y_d, \nonumber\\
(Y^{D}_{23})_{22} & = & \frac{y_s+y_b}{2} -(y_b-y_s)s_{23d}c_{23d}, \nonumber\\
(Y^{D}_{23})_{23} & = &\frac{y_b-y_s}{2} -(y_b-y_s)s_{23d}^2=(Y^{D}_{23})_{32}, \nonumber\\
(Y^{D}_{23})_{33} & = & \frac{y_s+y_b}{2} +(y_b-y_s)s_{23d}c_{23d} .\ 
\eea
As this Yukawa matrix must have the form in Eq.~\ref{23symbreak} we have
 $(Y^{D}_{23})_{23}=(Y^{D}_{23})_{33}$ which then leads to
 \bea
 \tan{\theta_{23d}} & = & \frac{1}{2} \left[ z_d-1 + \sqrt{z_d^2-6z_d+1} \right], \
 \label{theta23solnd}
 \eea
 where $z_d={ y_s \over y_b} =\pm{ m_s \over m_b}$ and we have chosen the solution that leads to small angle
 $\theta_{23d}$
 and hence to small flavor symmetry breaking.
 Keeping terms to first order in $z_d$ we get
 \bea
 \tan{\theta_{23d}} & \approx  & -z_d. \
 \label{theta23solnapproxd}
 \eea
  We further obtain for $\chi_d$ and $y_B$ in Eq.~\ref{23symbreak},
 \bea
 \chi_d &=& - \tan{2 \theta_{23d}} \approx 2z_d, \nonumber\\
 y_B & = & (y_b-y_s)\cos{ 2 \theta_{23d}}.\
 \label{chi}
 \eea
 The elements of $S^D_{23}$ are not related to those of $Y^D_{23}$. We only require that they share the same flavor symmetry. Assuming there is no breaking of flavor symmetry in $S^D_{23}$,
new FCNC contributions, involving $ b \to s$ transitions, are generated when we transform to  the mass basis. We have,
\bea
S^{D'}_{23} & = & R_{23d}^{\dagger} W^{d\dagger}_{23}S^D_{23} W^d_{23} R_{23d}. \
\eea
The  nonzero terms in $S^{D'}$ are,
\bea
(S^{D'}_{23})_{11} & = & q_{11}, \nonumber\\
(S^{D'}_{23})_{22} & = & \frac{1}{2} \left( q_{22}-q_{23} \cos{2 \theta_{23d}} \right), \nonumber\\
(S^{D'}_{23})_{23} & = &-\frac{1}{2}{q_{23}} \sin{ 2 \theta_{23d}}=(S^{D'}_{23})_{32}, \nonumber\\
(S^{D'}_{23})_{33} & = &\frac{1}{2} \left( q_{22}+q_{23} \cos{2 \theta_{23d}} \right)  .\ 
\eea
Hence there is FCNC  $b \to s$ transitions that is suppressed by 
$ \sim { m_s \over m_b}$ (using Eq.~\ref{theta23solnapproxd}).
This new $b \to s$ transition, to a very good approximation, has the same flavor suppression as the SM as
${ m_s \over m_b} \sim \lambda^2 \sim |V_{tb}||V_{ts}| $.  The elements $q_{(11,22,23)}$ are complex and will have new CP violating phases. Finally we note that all FCNC effects are proportional to
$q_{23}$ which is the difference of the second higg's diagonal coupling to the second and third generation quarks in the flavor symmetric limit.

It is possible to include symmetry breaking in $S^D_{23}$. Consider the structure,
\bea
S^D_{23} & = &
\pmatrix{q_{11} & 0
&  0  \cr  0  &  \frac{1}{2}{q_{22}}(1+ 2 \eta_d)  &  \frac{1}{2}{q_{23}}(1+ 2 \epsilon_d)\cr  0  & \frac{1}{2}{q_{23}}  & \frac{1}{2}{q_{22}}}. 
 \label{symbreakSD}
\eea
In the mass basis the non zero elements in $S^{D'}$ are,
\bea
(S^{D'}_{23})_{11} & = & q_{11}, \nonumber\\
(S^{D'}_{23})_{22} & = & \frac{1}{2} \left( q_{22}-q_{23} \cos{2 \theta_{23d}} \right)+ 
\frac{1}{2}\eta_d q_{22}(1+ \sin{2 \theta_{23d}}) 
-\frac{1}{2}\epsilon_d q_{23} \cos{2 \theta_{23d}}, \nonumber\\
(S^{D'}_{23})_{23} & = &-\frac{1}{2}{q_{23}} \sin{ 2 \theta_{23d}}
-\frac{1}{2}\eta_d q_{22} \cos{2 \theta_{23d}}
-\frac{1}{2}\epsilon_d q_{23}(1+ \sin{2 \theta_{23d}}), 
 \nonumber\\
(S^{D'}_{23})_{32} & = &-\frac{1}{2}{q_{23}} \sin{ 2 \theta_{23d}}
-\frac{1}{2}\eta_d q_{22} \cos{2 \theta_{23d}}
+\frac{1}{2}\epsilon_d q_{23}(1- \sin{2 \theta_{23d}}),
 \nonumber\\
(S^{D'}_{23})_{33} & = &\frac{1}{2} \left( q_{22}+q_{23} \cos{2 \theta_{23d}} \right) 
+\frac{1}{2}\eta_d q_{22}(1- \sin{2 \theta_{23d}}) 
+\frac{1}{2}\epsilon_d q_{23} \cos{2 \theta_{23d}} 
 .\ 
 \label{eta}
\eea
We can make a couple of observations here. If $ \epsilon_d=0$ then we can choose $ \eta_d $ to set the off-diagonal elements of $S^{D'}$, responsible for the new FCNC effects, to be zero. This is possible as
$(S^{D'}_{23})_{23}=(S^{D'}_{23})_{32}$.
 Moreover for $q_{22} \sim q_{23}$, no new FCNC effects can be achieved with $\eta_d \sim \chi_d$, where $\chi_d$  defined in Eq.~\ref{23symbreak} represents the flavor symmetry breaking in $Y^D_{23}$. However, we cannot choose a non zero $\epsilon_d$ to get rid of FCNC effects
in $S^{D'}$ as in this case $(S^{D'}_{23})_{23} \ne (S^{D'}_{23})_{32}$  . This follows
 from the fact that $\eta_d$ represents flavor symmetry breaking effect that is similar to the one in $ Y^D_{23}$ while $ \epsilon_d$ represents flavor symmetry breaking  that is different than in $Y^D_{23}$.
 According to the principle of SFS  $\eta_d$ and $\epsilon_d$ can at most be similar in size to $\chi_d$. Hence from Eq.~\ref{eta} one observes that new FCNC  $ b \to s$ transitions are suppressed by factors of the size of $\chi_d$. As was pointed out in Ref.~\cite{bsmixing}, flavor symmetry breaking in $S^D_{23}$ is necessary to generate a new phase in $B_s$ mixing.
 
 So far we have concentrated only on the down type quark sector. It is possible that the up type quark sector has the same structure of Yukawa matrices and the same pattern of symmetry breaking as the down type quark sector. We then have the Yukawa matrices in the up type quark sector as,
 \begin{eqnarray}
 Y^U_{23} &= &\pmatrix{y_u & 0 &  0 \cr 0 & \frac{1}{2}{y_T}(1+2\chi_u) & \frac{1}{2}{y_T }\cr 0
&\frac{1}{2}{y_T} & \frac{1}{2}{y_T}},\nonumber\\
 S^U_{23} & = &
\pmatrix{p_{11} & 0
&  0  \cr  0  &  \frac{1}{2}{p_{22}}  &  \frac{1}{2}{p_{23}}\cr  0  & \frac{1}{2}{p_{23}}  & \frac{1}{2}{p_{22}}}. 
 \
 \label{23symbreak1}
 \eea
 The matrix $Y^U_{23}$  is now diagonalized
by, $U_u= W^u_{23} R^u_{23}$ where $W^u_{23}=W^d_{23}$ and
\bea
R^u_{23} & = & \pmatrix{1 & 0
&  0  \cr  0  &  c_{23u}   &  s_{23u}\cr  0  &-s_{23u}  &
c_{23u}}, \nonumber\\ 
c_{23u} & = & \cos { \theta_{23u}} ; s_{23u}  =  \sin { \theta_{23u}}. \
\label{23u}
 \eea
 We further obtain
 \bea
 \tan{\theta_{23u}} & = & \frac{1}{2} \left[ z_u-1 + \sqrt{z_u^2-6z_u+1} \right], \
 \label{theta23solnu}
 \eea
 where $z_u={ y_c \over y_t} =\pm{ m_c \over m_t}$. Here $y_{c,t}$ are the diagonal charm and top quark Yukawa couplings defined in Eq.~\ref{mass_Yukawa} and $m_{c,t}$ are the charm and the top quark masses.
 Keeping terms to first order in $z_u$ we get
 \bea
 \tan{\theta_{23u}} & \approx  & -z_u. \
 \label{theta23solnapprox}
 \eea
  We also get for $\chi_u$ and $y_T$ in Eq.~\ref{23symbreak1},
 \bea
 \chi_u &=& - \tan{2 \theta_{23u}} \approx 2 z_u, \nonumber\\
 y_T & = & (y_t-y_c)\cos{ 2 \theta_{23u}}.\
 \label{chi_u}
 \eea
  The elements of the Yukawa coupling, $S^U_{23}$, in the mass basis have the form,
 \bea
S^{U'}_{23} & = & R_{23u}^{\dagger} W^{u\dagger}_{23}S^U_{23} W^u_{23} R_{23u}, \
\eea
 with
 \bea
(S^{U'}_{23})_{11} & = & p_{11}, \nonumber\\
(S^{U'}_{23})_{22} & = & \frac{1}{2} \left( p_{22}-p_{23} \cos{2 \theta_{23u}} \right), \nonumber\\
(S^{U'}_{23})_{23} & = &-\frac{1}{2}{p_{23}} \sin{ 2 \theta_{23u}}=(S^{U'}_{23})_{32}, \nonumber\\
(S^{U'}_{23})_{33} & = &\frac{1}{2} \left( p_{22}+p_{23} \cos{2 \theta_{23u}} \right)  .\ 
\eea
 One then has new FCNC  $ t \to c $ transitions that are suppressed by $ \sim {m_c \over m_t}$. As in the case of $S^D_{23}$ we can also consider the effect of symmetry breaking in $S^U_{23}$ but we will not discuss this as the analysis is similar to the one for down type quarks. 
 
 The CKM matrix is now obtained as
 \bea
 V_{CKM} & = & V_{uL}^{\dagger}V_{dL},\nonumber\\
         &= & R^{u \dagger}_{23} R^d_{23}, \nonumber\\
         &= &\pmatrix{1 & 0
&  0  \cr  0  &  \cos{\theta_{23}}   &  \sin{\theta_{23}}\cr  0  &-\sin{\theta_{23}}  &
\cos{\theta_{23}}}, \ 
 \label{ckm2gen}
 \eea
 where $ \theta_{23}= \theta_{23d}- \theta_{23u} $. This gives a  description of the quark mixing if we neglect mixing effects involving the first generation \cite{fxing}.
 
 To obtain a realistic CKM matrix, the mixing involving the first generation have to be taken into account.
 We will assume that the Yukawa matrices $Y^{U,D}_{23}$  are diagonalized by unitary matrices $U_d$  for down type quarks and $V_u$ for up type quarks with
 \bea
 U_d & = & W_{23}R^{d}_{23}R^{d}_{12}, \nonumber\\
 V_u & = & W_{23}R^{u}_{23}R^{u}_{12}, \
 \label{diagmat}
 \eea
 where
 \bea
 R_{12u(d)} & = & \pmatrix{c_{12u(d)} & s_{12u(d)}
&  0  \cr  -s_{12u(d)}  &  c_{12u(d)}   &  0\cr 0  &
0& 1}, \
\label{12ud}
\eea
and $R^{u,d}_{23}$ are given by Eq.~\ref{23d} and Eq.~\ref{23u}.
 The CP violation CKM phase, $ \phi$, can be included by modifying either $R^u_{23}$ or $R^d_{23}$. We will choose to modify $R^d_{23}$ and take it to be given by,
 \bea
R^d_{23} & = & \pmatrix{e^{-i \phi} & 0
&  0  \cr  0  &  c_{23d}   &  s_{23d}\cr  0  &-s_{23d}  &
c_{23d}}, \nonumber\\ 
c_{23d} & = & \cos { \theta_{23d}} ; s_{23d}  =  \sin { \theta_{23d}}. \
\label{23dm}
 \eea
This scheme of mass matrices was considered in Ref~\cite{fxing}. Our scheme is equivalent to the one in Ref~\cite{fxing} up to a unitary transformation by $W_{23}$.

It is interesting to consider the limit $\theta_{23u(d)}=0$. In particular we are interested in the symmetric limit where the Yukawa couplings of the two generations are equal in the up and down quark sector. The diagonal Yukawa matrices are proportional to the identity matrix in the mass basis and hence $\Yud=R_{12u(d)} Y^{U,D}_{diag}R_{12u(d)}^{T}$ are also proportional to the identity matrix. 
%This is because $R_{12u(d)}$ are orthogonal matrices and commute with %$\Yud$. 
The angles
$\theta_{12u(d)}$ are undetermined in this case. Realistic value for
$\theta_{12u(d)}$ arise when $y_c \ne y_u$ and $y_s \ne y_d$.
 
Returning to the general case, in this scenario, the CKM matrix is given by,
\bea
V_{CKM} & = & V_u^{\dagger}U_d= R^{u \dagger}_{12}R^{u\dagger}_{23}R^{d}_{23}R^{d}_{12}.\
\label{ckm}
\eea
Explicitly we have,
\bea
V_{CKM} & = & \pmatrix{c_{12u}c_{12d} e^{-i \phi} +
s_{12u}s_{12d}c_{23} &
  c_{12u}s_{12d} e^{-i \phi} - s_{12u}c_{12d} c_{23} &
                             -s_{12u} s_{23} \cr
s_{12u}c_{12d} e^{-i \phi} - c_{12u}s_{12d} c_{23} & 
s_{12u}s_{12d} e^{-i \phi} + c_{12u}c_{12d} c_{23} &
c_{12u} s_{23} \cr s_{23} s_{12d} & -s_{23} c_{12d} &
 c_{23}}. \nonumber\\
\label{ckmfull}
\eea
In the above, $s_{12d}, s_{12u} \sim \lambda$ and $s_{23}= \sin{\theta_{23}}$ (which is defined in Eq.~\ref{ckm2gen})$ \sim \lambda^2$. 
We will assume that in the limit $\theta_{12u(d)}=0$ the structure of the Yukawa couplings are given by Eq.~\ref{23symbreak} and Eq.~\ref{23symbreak1} for the down and up type quark sectors. In this limit $\theta_{23d}$ and $\theta_{23u}$ are given by Eq.~\ref{theta23solnd} and Eq.~\ref{theta23solnu}. These solutions produce $\theta_{23}$ consistent with experiments. Hence with  $\theta_{12u(d)} \ne 0$ we expect the values for $\theta_{23d}$ and $\theta_{23u}$ to be not very different from those in Eq.~\ref{theta23solnd} and Eq.~\ref{theta23solnu}.

 One can consider specific structures for the Yukawa couplings to obtain solutions
for $\theta_{12u(d)}$ and $\theta_{23u(d)}$ in terms of quark masses.
This is beyond the scope of the paper and we will not discuss it. 
We note that with $\theta_{23}=0$ there is no mixing involving the third generation.

We turn now to FCNC effects involving $S^{U,D}_{23}$. We will make
the important assumption that all new FCNC effects vanish if the symmetry breaking involving the second and third
generations are turned off. Hence, in our scheme there is no FCNC if $\theta_{23d}=0$ and $\theta_{23u}=0$. We will again concentrate on the down quark sector though  similar results apply to the up quark sector. 
New FCNC effects are generated by the off-diagonal elements of
$S^{D'}_{23}$ where,
\bea
 S^{D'}_{23} & = &U_{d}^{\dagger} S^D_{23} U_d. \
\label{sdfcnc}
\eea
The requirement that $S^{D'}_{23}$ is diagonal when $\theta_{23d}=0$ leads to the following condition
\bea
q_{11} = \frac{1}{2}( q_{22}- q_{23}). \
\label{condition}
\eea
This tells us ( see Eq.~\ref{diag23sym})  that in the flavor symmetric limit the second higgs has the same coupling to the first two generations.
We can now write the elements of $S^{D'}_{23}$ in Eq.~\ref{sdfcnc} as,
\bea
(S^{D'}_{23})_{11} & = & \frac{1}{2}q_{22}+ \frac{1}{2}q_{23} \left(
1-2c_{23d}^2-2c_{12d}^2+2c_{12d}^2c_{23d}^2 \right), \nonumber\\
(S^{D'}_{23})_{12} & = & -q_{23}s_{12d}c_{12d}s_{23d}^2 =(S^{D'}_{23})_{21}, \nonumber\\
(S^{D'}_{23})_{13} & = & q_{23}s_{12d}s_{23d}c_{23d}=(S^{D'}_{23})_{31}, \nonumber\\
(S^{D'}_{23})_{22} & = & \frac{1}{2}q_{22}- \frac{1}{2}q_{23} \left(
1-2c_{12d}^2+2c_{12d}^2c_{23d}^2 \right), \nonumber\\
(S^{D'}_{23})_{23} & = &-q_{23}c_{12d}s_{23d}c_{23d}=(S^{D'}_{23})_{32},
 \nonumber\\
(S^{D'}_{23})_{33} & = &\frac{1}{2}q_{22}+ \frac{1}{2}q_{23} \left(
2c_{23d}^2-1 \right) .\ 
 \label{fcnc3}
\eea
{}From the above we find that $ s \to d$ transitions are suppressed by $ \sim \lambda^5$, $ b \to d$ transitions are suppressed by $ \sim \lambda^3$ and 
$ b \to s $ transitions are suppressed by $ \sim \lambda^2$. A similar pattern of FCNC effects will occur in the top quark sector. 

Let us briefly discuss symmetry breaking in $S^D_{23}$. Let us assume the that flavor symmetry breaking has the same form as in Eq.~\ref{symbreakSD}.
In general, the symmetry breaking parameters $\eta_d$ and $\epsilon_d$ have to be $O(\lambda^4)$ to produce the $ s \to d$ transition that is suppressed by $\sim \lambda^5$. However if the symmetry breaking in the diagonal and non diagonal elements are the same, which happens with $
\eta_d q_{22}=\epsilon_d q_{23}$, then
$\eta_d$ and $\epsilon_d$ need be suppressed by only $O(\lambda^2)$ to produce the $ s \to d$ transition with suppression  $\sim \lambda^5$.

{\bf{ Leptonic Sector}} We now turn our attention to the leptonic sector. We will assume the same flavor symmetry limit and the same pattern of flavor symmetry breaking for the charged lepton sector as in the quark sector.
Hence in the flavor symmetry limit we have,
\bea
Y^L_{23} & = &
\pmatrix{l_{11} & 0
&  0  \cr  0  &  \frac{1}{2}{l_{22}}  &  \frac{1}{2}{l_{23}}\cr  0  
&\frac{1}{2}{l_{23}}  &\frac{1}{2}{l_{22}}}, 
 \nonumber\\
 S^L_{23} & = &
\pmatrix{r_{11} & 0
&  0  \cr  0  &  \frac{1}{2}{r_{22}}  &  \frac{1}{2}{r_{23}}\cr  0  & \frac{1}{2}{r_{23}}  & \frac{1}{2}{r_{22}}}. 
 \
 \label{23lsym}
 \eea

The Yukawa matrices  $Y^L_{23}$ and $S^L_{23}$ are diagonalized by the same unitary matrix $W^{l}_{23}=W_{23}$ given by,
\bea
W^l_{23} &  = & \pmatrix{1 & 0
&   0  \cr   0   &  -\frac{1}{\sqrt{2}}   &  \frac{1}{\sqrt{2}}\cr   0
&\frac{1}{\sqrt{2}} &  \frac{1}{\sqrt{2}}}.\
\label{wl} 
\eea 
Writing the diagonalized Yukawa matrices as $Y^{L}_{23diag}$ and
$S^L_{23diag}$ we have,
\bea 
Y^L_{23diag} & = & W^{l \dagger}_{23} Y^L_{23} W^l_{23} =\pmatrix{l_{11} & 0
&  0  \cr  0  &   \frac{1}{2}{(l_{22}-l_{23})}  &  0\cr  0  &0  &
\frac{1}{2}{(l_{22}+l_{23})}}, \nonumber\\ 
S^L_{23diag} & = & W^{l \dagger}_{23} S^L_{23} W^l_{23} =\pmatrix{r_{11} & 0
&  0  \cr  0  &   \frac{1}{2}{(r_{22}-r_{23})}  &  0\cr  0  &0  &
\frac{1}{2}{(r_{22}+r_{23})}}. \nonumber\\ 
\label{diag23symlepton} 
\eea 
After including flavor symmetry breaking, we will assume that the Yukawa matrix $Y^L_{23}$  is diagonalized by the unitary matrix $U_l$ which is
 given by, 
 \bea
 U_l & = & W_{23}R^{l}_{23}R^{l}_{12}, \
 \label{diagmatlepton}
 \eea
 where $R^{l}_{23}$ and $R^{l}_{12}$ have the same structure as in the quark sector.
 
 FCNC effects involving $S^{L}_{23}$ are represented by the off-diagonal elements in the mass basis. As in the quark sector we will make
the  assumption that 
 all new FCNC effects vanish if the symmetry breaking involving the second and third
generation charged leptons is turned off. Hence, in our scheme there is no FCNC if $\theta_{23l}=0$. 
Denoting $S^{L}_{23}$ in the mass basis as
$S^{L'}_{23}$ we have,
\bea
 S^{L'}_{23} & = &U_{l}^{\dagger} S^L_{23} U_l. \
\label{sdfcnclepton}
\eea
The requirement that $S^{L'}_{23}$ is diagonal when $\theta_{23l}=0$ leads to the following condition
\bea
r_{11} = \frac{1}{2}( r_{22}- r_{23}). \
\label{conditionlepton}
\eea
As in the quark sector, this tell us( from Eq.~\ref{diag23symlepton})  that in the flavor symmetric limit the second higgs has the same coupling to the first two generations of charged leptons.
We can now write the elements of $S^{L'}_{23}$ as,
\bea
(S^{L'}_{23})_{11} & = & \frac{1}{2}r_{22}+ \frac{1}{2}r_{23} \left(
1-2c_{23l}^2-2c_{12l}^2+2c_{12l}^2c_{23l}^2 \right), \nonumber\\
(S^{D'}_{23})_{12} & = & -r_{23}s_{12l}c_{12l}s_{23l}^2 =(S^{L'}_{23})_{21}, \nonumber\\
(S^{L'}_{23})_{13} & = & r_{23}s_{12l}s_{23l}c_{23l}=(S^{L'}_{23})_{31}, \nonumber\\
(S^{L'}_{23})_{22} & = & \frac{1}{2}r_{22}- \frac{1}{2}r_{23} \left(
1-2c_{12l}^2+2c_{12l}^2c_{23l}^2 \right), \nonumber\\
(S^{L'}_{23})_{23} & = &-r_{23}c_{12l}s_{23l}c_{23l}=(S^{L'}_{23})_{32},
 \nonumber\\
(S^{L'}_{23})_{33} & = &\frac{1}{2}r_{22}+ \frac{1}{2}r_{23} \left(
2c_{23l}^2-1 \right) .\ 
 \label{fcnc3lepton}
\eea
{}From the above we find that $ \mu \to e$ transitions are most suppressed followed by $ \tau \to e$ transitions  and 
$ \tau \to \mu $.

To get an idea of the size of the FCNC effects we choose the structure of $Y^L_{23}$ in the limit $\theta_{12l}=0$ to be, 
\begin{eqnarray}
 Y^L_{23} &= &\pmatrix{l_e & 0 &  0 \cr 0 & \frac{1}{2}l_{T}(1+2\chi_l) & \frac{1}{2}{l_{T} }\cr 0
&\frac{1}{2}{l_{T}} & \frac{1}{2}{l_{T}}}.\
 \label{23dsymbreak1}
 \eea
As in the quark sector we can solve for $\theta_{23l}$ as
\bea
 \tan{\theta_{23l}} & = & \frac{1}{2} \left[ z_l-1 + \sqrt{z_l^2-6z_l+1} \right], \
 \label{theta23solnlepton}
 \eea
 where $z_l ={ l_{\mu} \over l_{\tau}}= \pm { m_{\mu} \over m_{\tau}} $.
 Here $l_{\mu(\tau)}$ are the Yukawa couplings defined in Eq.~\ref{mass_Yukawa}.
 Keeping terms to first order in $z_l$ we get
 \bea
 \tan{\theta_{23l}} & \approx  & -z_l. \
 \label{theta23solnapproxlepton}
 \eea
  We also get for $\chi_l$ and $l_T$ in Eq.~\ref{23dsymbreak1},
 \bea
 \chi_l &=& - \tan{2 \theta_{23l}} \approx 2z_l, \nonumber\\
 l_{T} & = & (l_{\tau}-l_{\mu})\cos{ 2 \theta_{23l}}.\
 \label{chi_l}
 \eea
 where $l_{\mu(\tau)}$ are the Yukawa couplings of the muon and tau lepton defined in Eq.~\ref{mass_Yukawa}.

Assuming $ \theta_{23l}$ to be not very different from that in Eq.~\ref{theta23solnlepton} with $\theta_{12l}\ne 0$,  we find from  Eq.~\ref{fcnc3lepton} that FCNC $ \tau \to \mu$ transition is suppressed by 
$ \sim { m_{\mu} \over m_{\tau} }$, $ \tau \to e $ transition is suppressed by 
$ \sim { m_{\mu} \over m_{\tau} }s_{12l}$ and $ \mu \to e$ transition is suppressed by 
$ \sim ({ m_{\mu} \over m_{\tau} })^2 s_{12l}$.

We can obtain information about the size of $\theta_{12l}$ by considering deviations from the TBM mixing by including the symmetry breaking in the charged lepton sector. The leptonic mixing matrix is now given by
\bea
U_{PMNS}&= U^{\dagger}_\ell U_\nu, \
\label{nuleptonfinal}
\eea
where
$U_\ell  =  W_{23} R^l_{23}R^l_{12}$ and $U_{\nu}$ is given in Eq.~\ref{tbmfactor}. Discussion on the deviation from the TBM form can be found in Ref.~\cite{dattaTBM, devs}. In particular one finds \cite{dattaTBM}, 
\bea
s_{12l} &= & -  s,\
\eea
where
$ -0.11<s<0.04 $ \cite{king}.

It is interesting to note that $\tau \to \mu$ and $ b \to s$ transitions have similar size flavor suppression as $ {m_{\mu} \over m_{\tau}} \sim
{m_s \over m_b} $ if we choose $m_s \sim $ 200 MeV and $m_b \sim $ 4.2 GeV. However the rates for $ \tau \to \mu$ and $b \to s$ transitions depend on the phases in the Yukawa couplings $S^L_{23}$ and $S^D_{23}$ and can be very different.

\section{ Conclusion}
In conclusion we have considered in this paper a new scheme to suppress FCNC effects from beyond the SM physics. This idea is based on the principle of shared flavor symmetry. The principle states that there exists a flavor symmetric limit where the SM and NP sources of FCNC effects share the same flavor symmetry and there are no new FCNC contributions beyond those of the SM. In a specific extension of the SM- the two higgs doublet model-  the Yukawa couplings of both the higgs doublet are diagonalized by the same transformation in the flavor symmetric limit. The transformation matrix is composed of pure numbers. In the flavor symmetric limit the quark mixing matrix and the lepton mixing matrix have  fixed forms. Realistic mixing matrices are obtained from the breaking of the flavor symmetry. New FCNC effects are generated by non universal breaking of the flavor symmetry in the SM and NP sources of FCNC effects. The size of these FCNC effects are determined by the breaking of the flavor symmetry in the SM sector. Within the two higgs doublet model we showed how a realistic quark and lepton mixing matrix may arise 
and how new FCNC effects can be generated which are sufficiently suppressed to be consistent with experiments.

%\bigskip
%\noindent {\bf Acknowledgments}:
%\bigskip
%This work is financially supported by NSERC of Canada. We thank David %London, R.N. Mohapatra and Sandip Pakvasa for useful discussion.

%%%%%%%%%%%%%%%%%%%%% REFERENCES %%%%%%%%%%%%%%%%%%%%%%%%%%%%%%%%

\end{document}